\documentclass[9pt,twocolumn,twoside]{osajnl}
\journal{ol} % Choose journal (ao, aop, josaa, josab, ol, pr)

\setboolean{shortarticle}{true}
\title{Intracavity Brillouin gain characterization based on cavity ringdown spectroscopy}

\author[1]{Ananthu Sebastian}
\author[1,*]{Stéphane Trebaol}
\author[1]{Pascal Besnard}

\affil[1]{Univ Rennes, CNRS, Institut FOTON - UMR 6082, F-22305 Lannion, France}

\affil[*]{Corresponding author: stephane.trebaol@enssat.fr}
\usepackage{makecell}

\begin{abstract}
We report a technique based upon the cavity ringdown method that enables to characterize the Brillouin gain coefficient directly in a laser cavity. Material gain, optical cavity parameters and lasing properties can be extracted from measurements whithin a single experiment. 
\end{abstract}

\setboolean{displaycopyright}{true}

\begin{document}

\maketitle

\section{Introduction}
The constant need for laser spectral purity improvements is driven by the growing panel of applications in fundamental \cite{Jiang2011, Predehl2012} and applied physics \cite{Geng2005, Rodrigo2010}. One of the most promising approach to generate compact and narrow linewidth lasers is based on the stimulated Brillouin scattering (SBS) optical nonlinearity \cite{Ippen1972}. The establishement of SBS in an optical cavity \cite{Smith1991} gives rise to the coherent emission of a Stokes wave. Impressive noise performances have been reported in such Brillouin lasers \cite{Loh2016, Suh2017, Sebastian2018}. To reach such high laser performances, one of the most important parameters to evaluate is the material gain coefficient.\\
The brillouin gain coefficient, can be expressed in function as material parameters by \cite{Agrawal2000}:
\begin{equation}
	g_B=\frac{{2}{\pi}{n^7} {p^2_{12}}}{{c}{\rho_0}{\lambda^2_p}{\Delta \nu_B}{V_A}}
	\label{gain}
\end{equation}
where c is the vacuum velocity of light, $\lambda_p$ the laser pump wavelength and $\Delta\nu_B$ is the Full Width at Half Maximum (FWHM) of the Brillouin gain. Other parameter values are given in table \ref{table1}.\\
Usually, the $g_B$ may be estimated through classical pump-probe experiments \citep{Shibata1987}, self-heterodyne \citep{Tkach1986}, Fabry-Perot interferometry \cite{Ippen1972} or threshold power determination \cite{Cotter1982,Abedin2005}. In those methods, the SBS phenomenon is generated by injecting a pump power signal in a single-pass waveguide. Close to SBS threshold, a Stokes signal can be efficiently created in the counter propagating direction of the incident pump.\\
The Stokes frequency is downshifted from the pump frequency by the Stokes shift $\nu_B$. The associated gain profile is described by:
\begin{equation}
g_B(\nu)=g_B\frac{(\Delta\nu_B/2)^2}{(\nu-\nu_B)^2+(\Delta\nu_B/2)^2}
\label{SBS}
\end{equation}
Probing the gain profile by above mentionned methods allows to determine $\Delta\nu_B$. Then introducing $\Delta\nu_B$ value and material constants, extracted by other experimental means or from calculations, into Eq. (\ref{gain}) gives a Brillouin gain coefficient estimation. Those methods suffer from several drawbacks. First, to reach the SBS threshold, one have to either use long waveguide \cite{Cotter1982} or high pump power \cite{Abedin2005}, which are not always suitable depending on the material and cavity design (waveguide, microresonator). And secondly, the $g_B$ parameter can be inferred at the expense of knowing material constants, which are not systematically available depending on the material under study.\\ 
The cavity ringdown method (CRDM) has been successfully implemented in various optical cavities \cite{Dumeige2008,Conti2011,Henriet2015} to determine coupling regime and dispersive properties \cite{Dumeige2008}, selective amplification in erbium doped media \cite{Rasoloniaina2014}, mode coupling \cite{Trebaol2010} and nonlinear parameters \cite{Savchenkov2007,Rasoloniaina2015}.\\
\begin{figure*}[htbp]
\begin{center}
\includegraphics[width=\linewidth]{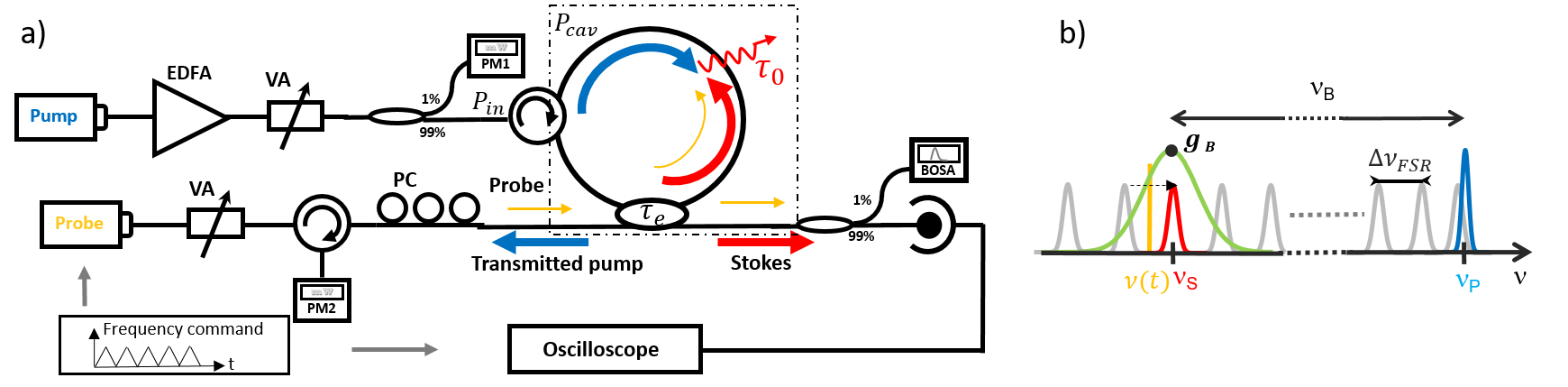}
\caption{a) Experimental setup for Brillouin gain cavity ringdown determination. EDFA: Erbium-doped fiber amplifier, VA: variable attenuator, PC: polarization controller, BOSA: Brillouin optical spectrum analyzer, PM : Powermeter. b) Spectral overview of the CRDM method. Pump laser line (blue), Brillouin gain curve (green), probed cavity mode (red) and probing laser line (yellow). $\nu_B$ corresponds to the Brillouin shift, where $\Delta\nu_{FSR}$ stands for the spectral spacing between cavity modes.}\label{setup}
\end{center}
\end{figure*}
In this letter, we propose to extend the use of the CRDM to retrieve the Brillouin gain coefficient. Performing Brillouin gain characterization inside the laser cavity allows to take advantage of the light recirculation of the Stokes wave inside the resonator. In fact, this tends to strongly reduce the length of the waveguide and the applied laser-pump power required to reach the SBS threshold in comparison to usual pump-probe techniques. Moreover, the gain coefficient is retrieved without the need of material parameter values. It allows also to identify several nonlinear processes that can occur for example in high-Q resonator as thermal drift and Kerr effect \cite{Rasoloniaina2015}.\\
The paper is organized as follows : First, we describe the procedure to retrieve the intracavity Brillouin gain coefficient from the CRDM technique. Then, we detail the experimental bench used to generate and probe the SBS gain in the cavity and finally we detail and discuss our results.
%\begin{figure}%[position]
%\includegraphics[width=\linewidth]{figures/spectra.png}
%\caption{False-color image, where each pixel is assigned to one of seven reference spectra.}\label{spectra}
%\end{figure}
\section{Principle}
\begin{figure}%[position]
\includegraphics[width=\linewidth]{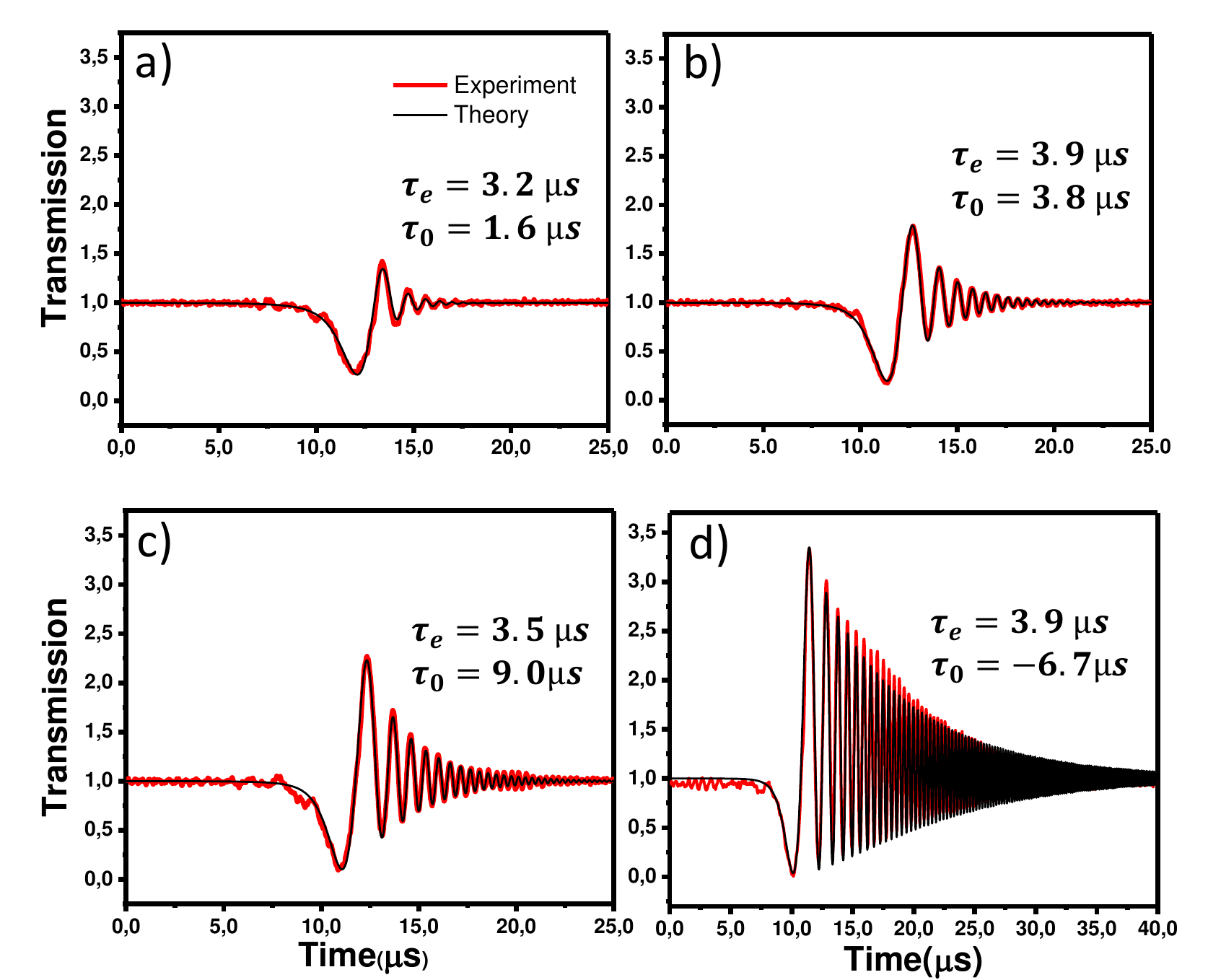}
\caption{Transient responses of the probed cavity mode for various laser pump powers. From a) to d) the resonator-coupling regime shifts with increasing pump power from under coupling ($P_{in}=13.8$ mW), critical coupling ($P_{in}=39.1$ mW), over coupling ($P_{in}=41.7$ mW) and selective amplification regime ($P_{in}=58.5$ mW).}\label{CRDM}
\end{figure}

Figure \ref{setup} b) recalls the spectral arrangement of waves taking part in the SBS process when studied in a resonant cavity. The blue line corresponds to the pump signal at frequency $\nu_P$ that produces a Stokes gain (green curve, Eq. (\ref{SBS})) through the SBS process. In the background of the figure, the periodic distribution of the ring cavity modes are colored in gray. In our configuration, the pump signal is not resonant within a cavity round-trip. Nevertheless, the perimeter of the resonator, that fixed the free spectral range of the cavity ($\Delta \nu_\text{FSR}$), and the spectral gain distribution insure to favor the amplification of one cavity mode (red color line), that we call the Stokes-mode. We consider that the Stokes-mode at the frequency $\nu_S$ is seeded at the maximum of the gain value $g_B$.\\
We have now to determine how the Stokes-mode linear-losses might be compensated by the Brillouin gain as a function of the incident pump power $P_{in}$. The following model description is based on the coupled mode theory described elsewhere \cite{Haus1984, Dumeige2008} and adapted to the present study for the purpose of Brillouin gain characterization.\\
The resonator schematic is shown in Fig. \ref{setup} a) (dashed square). We consider the temporal amplitude evolution $u_\text{S}(t)$ of the Stokes-mode. Its total photon lifetime $\tau/2$ experienced by this cavity mode is:
\begin{equation}
\frac{1}{\tau}=\frac{1}{\tau_0}+\frac{1}{\tau_e}
\end{equation}
where $\tau_0/2$ and $\tau_e/2$ are the intrinsic photon and the coupler lifetimes respectively. The intrinsic photon lifetime expresses the losses or gain of the cavity while the coupler lifetime relates to the coupling strength between the cavity and the input/output fiber. We can relate the coupling coefficient $|\kappa^2|$ and the intensity round-trip attenuation $a^2$ to their respective photon lifetime. Indeed, in the high finesse approximation ($|\kappa|^2 \times a^2 \approx 1$)\cite{Yariv2000}, the coupler lifetime can be related to the coupling coefficient by $|\kappa^2|=2\tau_L/\tau_e$ where $\tau_L$ is the photon round-trip time. The intrinsic photon lifetime $\tau_0$ is related to the intensity round trip attenuation $a^2$ by:
\begin{equation}
a^2=1-2\tau_L/\tau_0
\label{round_trip_intensity}
\end{equation}
Then a positive $\tau_0$ corresponds to an optical attenuation with $a^2<1$ while a negative $\tau_0$ implies optical Brillouin gain with $a^2>1$. Moreover, the intensity round-trip attenuation of the Stokes-mode can be expressed as:
\begin{equation}
a^2=\beta\times e^{-\alpha_L L} e^{g_B P_{cav}L_\text{eff}/A_\text{eff}}=a^2_\text{op}e^{g_B P_{cav}L_\text{eff}/A_\text{eff}}
\label{asquare}
\end{equation}
where $\beta$ is the inner local losses including the contribution of splices, circulator and coupler; $\alpha_L$ the fiber-loss coefficient; $A_\text{eff}$ the effective fiber-mode area. $L$ and $L_\text{eff}$ are respectively the cavity length and the effective interaction length. $a^2_\text{op}$ is the intensity round-trip attenuation of the cold cavity mode without stimulated Brillouin scattering. Here $P_{cav}$ is the nonresonant intracavity pump power.\\
As can be seen in Eq. (\ref{asquare}), increasing the pump intensity $P_{cav}$, allows the resonator linear losses experienced by the Stokes-mode to be compensated over a single round-trip. Thereby, intensity round-trip attenuation $a^2$ and intrinsic photon lifetime $\tau_0$ can be tuned through pump intensity keeping constant the coupler lifetime.\\ 
$\tau_0$ can be evaluated experimentally using the CRDM technique and therefore $a_0^2$ and $a^2$. Indeed, the probe signal coupled in the Stokes mode, through the coupler, will experience attenuation or amplification within a round-trip. It follows that, experimental estimation of the Brillouin gain coefficient can be obtained by using Eq. (\ref{round_trip_intensity}) together with Eq. (\ref{asquare}):
\begin{equation}
g_B=\frac{A_{eff}}{P_{cav}L_\text{eff}}\text{ln}\frac{a^2}{a^2_\text{op}}
\label{g0}
\end{equation}
The CRDM technique consists in probing the Stokes-mode by sweeping the frequency of a tunable narrow linewidth laser (yellow color line on Fig. \ref{setup}) across the resonance. The sweeping speed of the laser is tuned sufficiently fast to observed ringing phenomenom characteristic from the transient response of the cavity. In its current version, our experimental bench allows to study resonator Q-factor as low as $10^7$. The ringing effect is a signature of interferences between the probing laser and the cavity mode waves at the output of the coupler. In a previous paper \cite{Dumeige2008}, we have determined an analytical expression for the transient response of a resonator as function of $\tau_0$, $\tau_e$ and the sweeping speed. By use of a least square method, this expression can be used to retrieve the resonator lifetimes and then the Brillouin gain coefficient as shown on Fig. \ref{CRDM}.
\section{Experimental setup}
The experimental setup is presented in Fig. \ref{setup} a). The fiber cavity is composed of a $L=20~m$ polarization maintaining silica fiber distributed as $L_1=19$ m  from coupler to isolator and $L_2=1$ m between isolator and coupler. Fiber length uncertainty is of the order of $\pm 0.6~m$. This cavity length giving a free spectral range (FSR) of $\Delta\nu_{\text{FSR}}=10~\pm0.3 ~\text{MHz}$. The transient response acquired by the CRDM technique results from the beating between the transmitted probe and the output coupled Stokes-mode at the output of the coupler. The intensity of the probe, seeded in the Stokes-mode, has to be precisely determined after one round-trip. It is related to the input probe at the coupler position by the intensity round-trip attenuation (see Eq. (\ref{asquare})). The effective interaction length have to be precisely determined to optimized the evaluation of $g_B$. For our cavity configuration, $L_\text{eff}= L_1+ |t^2|\times L_2 = 19.9 \pm 0.6~m$ where the coupler transmission coefficient $|t^2|=(1-|k^2|)$ is introduced to incorporate the pump intensity attenuation through the coupler. The effective area of the fiber is $A_\text{eff}= 84.9~\pm 0.2~\mu$m$^2$. A part of the Stokes wave is extracted from the resonator through a 95/5 coupler.\\
Let's first describe the SBS setup. The laser pump intensity of a single mode laser with 250 kHz linewidth centered at 1550 nm, is first amplified through an EDFA and then adjusted by a variable attenuator. 1 $\%$ of the pump is extracted through a coupler for intensity power monitoring (PM1). The nonresonant intracavity pump power  $P_{cav}$ is determined just after the circulator and can be related to $P_{in}$ by $P_{cav}=\alpha\times P_{in}$ where $\alpha=0.758$ is the relative losses experimentally evaluated. The circulator ensures a single pass of the pump signal within the fiber loop preventing re-injection. For sufficiently intense pump intensity, stimulated Brillouin Stokes signal is generated in the opposite direction of the pump signal. Extracted Stokes signal is analyzed spectrally through a high resolution BOSA and the intensity is recorded by a photodiode. The lasing threshold is determined when pump depletion starts corresponding to a pump power $P_{th}$ equal to $62~\text{mW}$.\\
We will now describe the CRDM setup. All the measurements are performed for a pump power $P_{in}$ lower than the pump power lasing threshold $P_{th}$ (corresponding to a range of 1-100 mW in our main set-up) to be within the validy domain of the technique \cite{Dumeige2008}. The probe signal is provided from a tunable laser with 250 kHz linewidth. Before entering the cavity, the signal intensity is controlled by a variable attenuator (VA) and its polarization adjusted through polarization controller (PC). The probe laser frequency is positioned close to the Stokes-mode with the help of the 10 MHz-resolution BOSA. The probe frequency is continuously swept across the resonant Stokes signal by a frequency command allowing the sweeping speed and the scanned frequency range to be adjusted. The transmission of the cavity mode is then collected by a photodiode and observed on an oscilloscope triggered by the frequency command.  
\section{results}
\begin{figure}%[position]
\includegraphics[width=\linewidth]{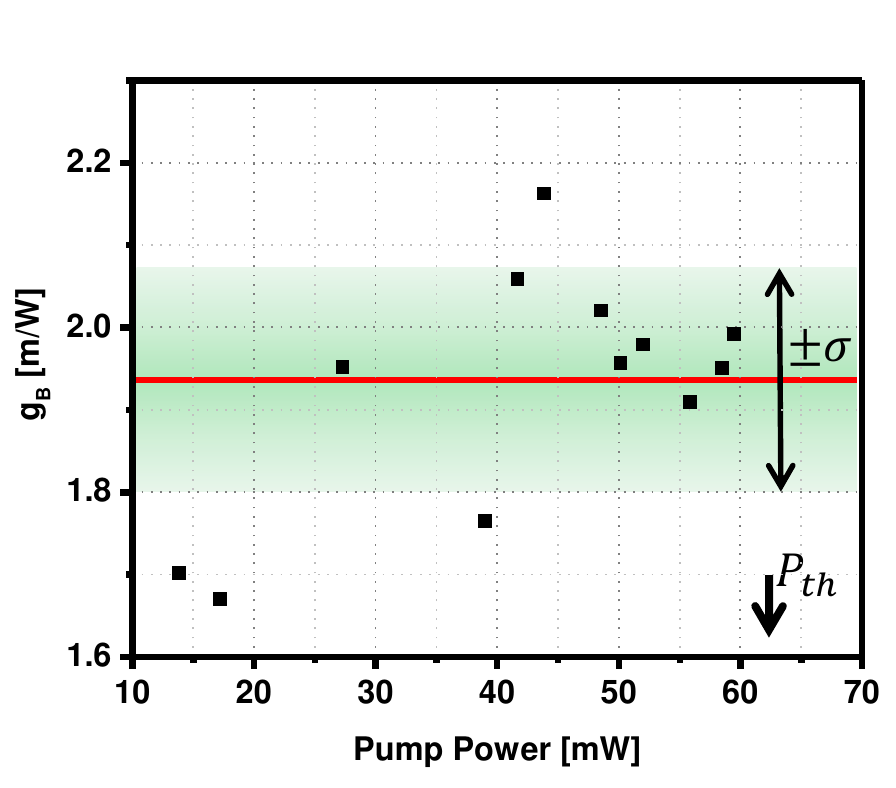}
\caption{Brillouin gain coefficient extracted from the CRDM signal for various input pump power. The mean value is equal to $g_B=1.94 \times 10^{-11} \pm 1.5 \times 10^{-12} m/W $. This measure of $g_B$ value is compared to other works in Tab. \ref{table}}\label{gB}
\end{figure}

CRDM signals are collected for input pump powers varying from 0 to $P_{th}$. Figure \ref{CRDM} presents various CRDM signals collected for pump power ranging from 13.8 to 58.5 mW (red color curves). The normalized transmission is plotted as function of time. For increasing pump power, the amplitude, the number and the frequency of oscillations increase. To get more insight on those specific temporal signatures, data are fitted using the procedure described above and allows to determine $\tau_0$ and $\tau_e$ as function of $P_{in}$ (See Fig. \ref{CRDM}). As expected the $\tau_e$ value is constant over all the measurements because the coupler losses are fixed. For $P_{in}=0$ mW, the cold cavity parameters are extracted and gives $\tau_e=3.8$ $\mu$s and $\tau_0 = 1.2$ $\mu$s. Using the definition $Q=2\pi\nu_S\tau/2$ gives an estimated quality factor of $5.6 \times 10^8$. For $P_{in}= 13.8$ to $50.1$ mW, $\tau_0$ increases, modifying the coupling regime from undercoupling (cold cavity) to the critical coupling, the overcoupling and finally the transparency regime. For $50.1$ mW$ <P_{in}<P_{th}$, $\tau_0$ takes a negative value corresponding to the amplification regime. Indeed, the variation of $\tau_0$ as function of the pump power  refers to a progressive compensation of the linear losses until the amplification regime.\\
\begin{table}[htbp]
	\centering
	\caption{\bf Comparison of $g_B$ values obtained for single mode silica fiber \cite{Corning} in various works. SH stands for Self-Heterodyne.}
	\begin{tabular}{ccc}
		\hline
		Methods & $g_B$ value  & standard deviation \\
		& [$\times 10^{-11}$m$/$W] & [$\pm 10^{-12}$m$/$W]\\
		\hline
CRDM [our work] & $1.94 $ & $1.5$ \\
SH [silica value]  & $2.45$  & $1.8$ \\
SH [3$\%$ $GeO_2$ value] & $1.92$ & $1.4$ \\
Pump-probe \cite{Dragic2011} &  $2.29$ & - \\

      \hline
       \end{tabular} \label{table}
        \end{table}
\begin{table}[htbp]
	\centering
	\caption{\bf Material values for pure and 3$\%$ $GeO_2$ doped silica fibers.}
	\begin{tabular}{ccc}
		\hline
		Parameters & $SiO_2$  & $SiO_2 + 3\% GeO_2 $ \\
		& \cite{Dragic2011} & \cite{Lagakos1980}\\
		\hline
		Refractive index [n] &1.45 & 1.46 \\
		Elasto-optic coefficient $[p_{12}]$ & 0.271  & 0.236 \\
	    Density $[\rho (kg.m^-2)]$ & 2200 & 2244 \\
		\hline
	\end{tabular} \label{table1}
\end{table}    
From those fitted datas, we are able to retrieve $a^2$ (Eq. (\ref{asquare})) and then the value of the Brillouin gain $g_B$ by use of Eq. (\ref{g0}). This procedure is applied for each pump power giving estimations of the Brillouin gain parameter. Figure \ref{gB} summarized the extracted gain values as function of the pump power. The mean value (red dashed line) is equal to $1.94 \times 10^{-11}$m$/$W with a standard deviation (shaded green region) of $\pm 1.5\cdot 10^{-12}$ m$/$W. The uncertainty contribution to $g_B$ related to $A_\text{eff}$, $L_\text{B}$ and $P_{cav}$ gives  $0.6\times 10^{-12}$ m$/$W and do not enlarge the experimentally determined standard deviation of $g_B$. We can then conclude that the Brillouin gain standard deviation is related to the experimental extraction of $\tau_0$.\\
To evaluate the pertinence of the method, we compare our CRDM technique to usual self-heterodyne method \citep{Tkach1986}. 
The Brillouin gain bandwidth is experimentally estimated to $\Delta\nu_B= 27.5\pm 2$ MHz. Then, using Eq. (\ref{gain}), gain coefficient can be determined. As mentioned before, this determination depends on material parameters of the used fiber. Fabrication process conditions (doping, temperature, etc.) affect those parameters. For example, a commercial single mode fiber in the C band region has usually a $GeO_2$-doping concentration of few percents but this value is usually unknown precisely. We report in table \ref{table1} material values for pure silica and $3\%$-doped $GeO_2$ silica fiber. Using those parameters, Brillouin gain estimation gives $2.45$ and $1.92 \times 10^{-11}$m$/$W (see Table \ref{table}) respectively. The standard deviation, related to $\Delta\nu_B$ determination, is bounded between $\pm 1.4 \times 10^{-12}$ m$/$W$< \sigma<1.8 \times 10^{-12}$ m$/$W. Our $g_B$ determination is very close from the one obtained by self-heterodyne using the parameters of a commercial monomode fiber $3\%$-doped with $GeO_2$. Upon this example, we see that the material parameters strongly impact the estimation of the Brillouin gain. Comparison to other works on similar silica fibers (see Tab. \ref{table}) give comparable $g_B$ values.

\section{Conclusion}
A cavity ringdown method is applied to stimulated Brillouin scattering in fiber cavity. Contrary to usual techniques, this new method gives access, through a simple and single measurement, as well to the Brillouin gain coefficient of the fiber material as to the cavity parameters. We have shown that the fast sweeping ringdown technique allows Brillouin gain to be characterized. The proof of concept has been experimentally demonstrated with a silica fiber ring resonator. This allowed us to determine unambiguously the coupling regime and to estimate the Brillouin gain coefficient of the material composing the resonator without the needs of material constants knowledge. The comparison to usual pump-probe techniques gives good agreements.\\
This cavity ringdown method can be applied to any kind of resonators allowing for example, the determination of Brillouin gain coefficient in microresonators, exotic material fiber rings and whispering gallery mode resonators.

The present work is supported under project FUI AAP20 SOLBO, with the help of BPI FRANCE and P\^{o}le Images $\&$ R\'{e}seaux. We thank also UBL for its financial support.

The authors would like to thank P. Féron and Y. Dumeige for fruitfull discussions and for their careful reading of the manuscript.

%\section{References}
%
%Note that \emph{Optics Letters} uses an abbreviated reference style. Citations to journal articles should omit the article title and final page number; this abbreviated reference style is produced automatically when the \emph{Optics Letters} journal option is selected in the template, if you are using a .bib file for your references.
%
%However, full references (to aid the editor and reviewers) must be included as well on a fifth informational page that will not count against page length; again this will be produced automatically if you are using a .bib file.

% Bibliography
\bibliography{OL_CRDM}

\begin{thebibliography}{10}
\newcommand{\enquote}[1]{``#1''}

\bibitem{Jiang2011}
Y.~Jiang, A.~Ludlow, N.~D. Lemke, R.~W. Fox, J.~A. Sherman, L.-S. Ma, and C.~W.
  Oates, {\protect\JournalTitle{Nature Photonics}} \textbf{5}, 158 (2011).

\bibitem{Predehl2012}
K.~Predehl, G.~Grosche, S.~Raupach, S.~Droste, O.~Terra, J.~Alnis, T.~Legero,
  T.~H{\"a}nsch, T.~Udem, R.~Holzwarth \emph{et~al.},
  {\protect\JournalTitle{Science}} \textbf{336}, 441 (2012).

\bibitem{Geng2005}
J.~Geng, C.~Spiegelberg, and S.~Jiang, {\protect\JournalTitle{IEEE Photonics
  Technology Letters}} \textbf{17}, 1827 (2005).

\bibitem{Rodrigo2010}
P.~J. Rodrigo and C.~Pedersen, {\protect\JournalTitle{Optics Express}}
  \textbf{18}, 5320 (2010).

\bibitem{Ippen1972}
E.~Ippen and R.~Stolen, {\protect\JournalTitle{Applied Physics Letters}}
  \textbf{21}, 539 (1972).

\bibitem{Smith1991}
S.~Smith, F.~Zarinetchi, and S.~Ezekiel, {\protect\JournalTitle{Optics
  letters}} \textbf{16}, 393 (1991).

\bibitem{Loh2016}
W.~Loh, J.~Becker, D.~C. Cole, A.~Coillet, F.~N. Baynes, S.~B. Papp, and S.~A.
  Diddams, {\protect\JournalTitle{New Journal of Physics}} \textbf{18}, 045001
  (2016).

\bibitem{Suh2017}
M.-G. Suh, Q.-F. Yang, and K.~J. Vahala, {\protect\JournalTitle{Physical review
  letters}} \textbf{119}, 143901 (2017).

\bibitem{Sebastian2018}
A.~Sebastian, I.~V. Balakireva, S.~Fresnel, S.~Trebaol, and P.~Besnard,
  {\protect\JournalTitle{Opt. Express}} \textbf{26}, 33700 (2018).

\bibitem{Agrawal2000}
G.~P. Agrawal, \enquote{Nonlinear fiber optics,} in \emph{Nonlinear Science at
  the Dawn of the 21st Century,}  (Springer, 2000), pp. 195--211.

\bibitem{Shibata1987}
N.~Shibata, R.~G. Waarts, and R.~P. Braun, {\protect\JournalTitle{Optics
  letters}} \textbf{12}, 269 (1987).

\bibitem{Tkach1986}
R.~Tkach, A.~Chraplyvy, and R.~Derosier, {\protect\JournalTitle{Electronics
  Letters}} \textbf{22}, 1011 (1986).

\bibitem{Cotter1982}
D.~Cotter, {\protect\JournalTitle{Electronics Letters}} \textbf{18}, 495
  (1982).

\bibitem{Abedin2005}
K.~S. Abedin, {\protect\JournalTitle{Optics Express}} \textbf{13}, 10266
  (2005).

\bibitem{Dumeige2008}
Y.~Dumeige, S.~Trebaol, L.~Ghi{\c{s}}a, T.~K.~N. Nguyen, H.~Tavernier, and
  P.~F{\'e}ron, {\protect\JournalTitle{JOSA B}} \textbf{25}, 2073 (2008).

\bibitem{Conti2011}
G.~N. Conti, S.~Berneschi, F.~Cosi, S.~Pelli, S.~Soria, G.~C. Righini,
  M.~Dispenza, and A.~Secchi, {\protect\JournalTitle{Opt. Express}}
  \textbf{19}, 3651 (2011).

\bibitem{Henriet2015}
R.~Henriet, G.~Lin, A.~Coillet, M.~Jacquot, L.~Furfaro, L.~Larger, and Y.~K.
  Chembo, {\protect\JournalTitle{Optics letters}} \textbf{40}, 1567 (2015).

\bibitem{Rasoloniaina2014}
A.~Rasoloniaina, V.~Huet, T.~K.~N. Nguyen, E.~Le~Cren, M.~Mortier, L.~Michely,
  Y.~Dumeige, and P.~F{\'e}ron, {\protect\JournalTitle{Scientific Reports}}
  \textbf{4}, 4023 (2014).

\bibitem{Trebaol2010}
S.~Trebaol, Y.~Dumeige, and P.~F\'{e}ron, {\protect\JournalTitle{Physical
  Review A}} \textbf{81}, 43828 (2010).

\bibitem{Savchenkov2007}
A.~A. Savchenkov, A.~B. Matsko, M.~Mohageg, and L.~Maleki,
  {\protect\JournalTitle{Opt. Lett.}} \textbf{32}, 497 (2007).

\bibitem{Rasoloniaina2015}
A.~Rasoloniaina, V.~Huet, M.~Thual, S.~Balac, P.~F{\'e}ron, and Y.~Dumeige,
  {\protect\JournalTitle{JOSA B}} \textbf{32}, 370 (2015).

\bibitem{Haus1984}
H.~A. Haus, \emph{{Waves and fields in optoelectronics}} (Prentice-Hall, 1984).

\bibitem{Yariv2000}
A.~Yariv, {\protect\JournalTitle{Electronics Letters}} \textbf{36}, 321 (2000).

\bibitem{Corning}
\enquote{Corning smf-28 ultra optical fiber,}
  \url{https://www.corning.com/media/worldwide/coc/documents/Fiber/SMF-28%20Ultra.pdf},
  note = {Accessed 24 May 2019}.

\bibitem{Dragic2011}
P.~D. Dragic, {\protect\JournalTitle{Journal of Lightwave Technology}}
  \textbf{29}, 967 (2011).

\bibitem{Lagakos1980}
N.~Lagakos, J.~A. Bucaro, and R.~Hughes, {\protect\JournalTitle{Applied
  Optics}} \textbf{19}, 3668 (1980).

\end{thebibliography}

% Full bibliography added automatically for Optics Letters submissions; the following line will simply be ignored if submitting to other journals.
% Note that this extra page will not count against page length
%\bibliographyfullrefs{OL_CRDM}

%Manual citation list
%\begin{thebibliography}{1}
%\bibitem{Zhang:14}
%Y.~Zhang, S.~Qiao, L.~Sun, Q.~W. Shi, W.~Huang, %L.~Li, and Z.~Yang,
 % \enquote{Photoinduced active terahertz metamaterials with nanostructured
  %vanadium dioxide film deposited by sol-gel method,} Opt. Express \textbf{22},
  %11070--11078 (2014).
%\end{thebibliography}

\end{document}